\documentclass[prr,aps,twocolumn,reprint,superscriptaddress, floatfix]{revtex4-2}
\usepackage{amsmath}
\usepackage{amssymb}
\usepackage{graphicx}
\usepackage{color}
\usepackage{eurosym}
\usepackage[utf8]{inputenc}
\usepackage[T1]{fontenc}
\usepackage{multirow}
\usepackage{xcolor}
\usepackage{hyperref}

\hypersetup{colorlinks, linkcolor=blue, citecolor=blue, urlcolor=blue}

\begin{document}

\title{Stable supersolids and boselets in spin-orbit-coupled Bose-Einstein
condensates with three-body interactions}
\author{Rajamanickam Ravisankar}
\affiliation{Department of Physics, Zhejiang Normal University, Jinhua
321004, PR China}
\affiliation{Zhejiang Institute of Photoelectronics \&
Zhejiang Institute for Advanced Light Source, Zhejiang Normal University,
Jinhua, Zhejiang 321004, China}

\author{Sanu Kumar Gangwar}
\affiliation{Department of Physics, Indian Institute of Technology, Guwahati
781039, Assam, India}

\author{Henrique Fabrelli}
\affiliation{Centro Brasileiro de Pesquisas Físicas, 22290-180 Rio de
Janeiro, RJ, Brazil}

\author{Yongping Zhang}
\affiliation{Institute for Quantum Science and Technology, Department of
Physics, Shanghai University, Shanghai 200444, PR China}

\author{Paulsamy Muruganandam}
\affiliation{Department of Physics, Bharathidasan University,
Tiruchirappalli 620024, Tamil Nadu, India}
\affiliation{Department of
Medical Physics, Bharathidasan University, Tiruchirappalli 620024, Tamil
Nadu, India}

\author{Pankaj Kumar Mishra}
\affiliation{Department of Physics, Indian Institute of Technology, Guwahati
781039, Assam, India}

\author{Emmanuel Kengne}
\affiliation{School of Physics and Electronic Information Engineering,
Zhejiang Normal University, Jinhua 321004, China}
\author{Gao Xianlong}
\email{Corresponding author: gaoxl@zjnu.edu.cn}
\affiliation{Department of Physics, Zhejiang Normal University, Jinhua
321004, PR China}
\author{Boris A. Malomed}
\affiliation{Department of Physical Electronics, School of Electrical and Computer Engineering,
Faculty of Engineering, Tel Aviv University, P.O.B. 39040, Ramat Aviv, Tel Aviv, Israel, Israel}

\begin{abstract}
We explore the stability of supersolid striped waves, plane-wave boselets,
and other extended states in one-dimensional spin-orbit-coupled
Bose-Einstein condensates with repulsive three-body interactions (R3BIs),
modeled by quintic terms in the framework of the corresponding
Gross-Pitaevskii equations. In the absence of R3BIs, the extended states are
susceptible to the modulational instability (MI) induced by the cubic
attractive nonlinearity. Using the linearized Bogoliubov-de-Gennes
equations, we identify multiple new types of MI, including baseband,
passband, mixedband, and zero-wavenumber-gain ones, which give rise to %
deterministic rogue waves and complex nonlinear wave
patterns. Our analysis reveals that R3BIs eliminate baseband and
zero-wavenumber-gain MIs, forming, instead, phonon modes that enable stable
boselets. Additionally, mixedband and passband MIs are suppressed, which
results in a lattice-like phonon-roton mode that supports a stable
supersolid phase. These stable supersolids can be realized using currently
available ultracold experimental setup.
\end{abstract}

\maketitle

%%%%%%%%%%%%%%%%%%%%%%%%%%%%%%%%%%%%%%%%%%%%%%%%%%%%%%%%%%%%%%%%%%%%

\section{Introduction}

%%%%%%%%%%%%%%%%%%%%%%%%%%%%%%%%%%%%%%%%%%%%%%%%%%%%%%%%%%%%%%%%%%%%
Spinor Bose-Einstein condensates (BECs) serve as a versatile platform for
examining stable and unstable dynamics of matter-waves under the action of
intra- and intercomponent interactions~\cite{Miesner1999, Salasnich2003,
Kasamatsu2004, Kevrekidis2004,Kasamatsu2006}. In this context, the
introduction of synthetic spin-orbit coupling (SOC) in binary BECs~\cite%
{Lin2011} has sparked interest in the creation of novel stable phases of
quantum matter, particularly the stripe phase characterized by spontaneous
breaking of the gauge and translational symmetries \cite{Martone2013,
Li2013, Liao2018, Geier2021, JRLi2017, Putra2020, Geier2023}, leading to
manifestations of superfluidity and crystalline properties, ultimately
resulting in the emergence of the supersolid phase \cite{supersolid1,
supersolid2, supersolid3, supersolid4, supersolid5}. These phases have been
observed in solid helium~\cite{Kim2004, helium1, helium2} and dipolar BEC~%
\cite{dipolar-supersolid}, as well as in SOC bosonic condensates~\cite%
{JRLi2017, Putra2020, Geier2023}. However, in the presence of SOC, the
supersolid phase is unstable even in the presence of a trapping potential~%
\cite{Geier2023}, presenting a challenge to maintain stability under the
action of attractive two-body interactions (A2BIs). The stability analysis
is usually performed within the framework of linearized Bogoliubov-de-Gennes
(BdG) equations for small perturbations added to the stationary state \cite%
{Bogolyubov19471, Bogolyubov19472, Bogolyubov19473, Goldstein1997, Wang2010,
Wu2018, Zhang2012}.

Modulational instability (MI) is a key factor in understanding the nonlinear
dynamics of matter-waves, such as soliton trains and domain patterns~\cite%
{Miesner1999, Salasnich2003, Kasamatsu2004, Kevrekidis2004, Kasamatsu2006,
Hulet, Robinson, SK}, particularly in binary BECs with SOC~\cite{Bhat2015,
Porsezian, Kasamatsu, Ravisankar}. Besides matter-waves \cite{solitons1,
solitons2, solitons3, matterwaves1, Salasnich2003, matterwaves2,
matterwaves3, matterwaves4}, the MI analysis has significant implications
across various fields~\cite{ZakhOstr}, including nonlinear optics \cite%
{optics1, optics2, optics3, optics4, optics5, optics6, optics7, optics8},
photonics \cite{Skryabin1, Skryabin2}, liquid crystals~\cite{liquidcrystal},
fluids~\cite{fluids, McLean, Deconninck}, plasmas~\cite{electric, electric2}
and other discrete systems \cite{KP}. In this context, it has been reported
\cite{Baronio2014,Liu2023} that baseband modulational instability (BBMI) and
zero-momentum-gain MI are primarily responsible for the formation of rogue
waves (RWs). Additionally passband modulational instability (PBMI)
contributes to driving unstable nonlinear dynamics.

On the other hand, there has been significant interest that has emerged
among the ultracold community to realize bright \cite{solitontrain-exp1,
brightsoliton-exp2, Thompson} and dark~\cite{darksoliton-exp1, solitons3}
solitons of BECs in the laboratory experiment. In particular, bright-soliton
trains are known to be generated through MI from a plane-wave (PW) input,
which has been experimentally demonstrated to exist in the
quasi-one-dimensional BEC\ in the setup of $^{7}$Li atomic gas \cite%
{solitontrain-exp1,Hulet}, where the atomic interactions were tuned to be
attractive using the Feshbach-resonance technique~\cite{Feshbach1,
Feshbach2, Feshbach3}.
Characteristics of MI generated by the
nonlinear evolution of MI of plane waves have been realized experimentally  in
fiber optics~\cite{Kraych2019}. In BEC models that incorporate SOC and
attractive cubic terms in the Gross-Pitaevskii equations (GPEs), stable
one-dimensional (1D) solitons exhibit a smooth or striped inner structure
within a specific range of chemical potential under trapping but become
unstable without trapping~\cite{Achilleos2013-bs}, indicating a
characteristic related to MI.

It has been shown that the inclusion of quartic self-repulsive terms in the
GP equations, accounting for the Lee-Huang-Yang (LHY) correction to
mean-field theory, stabilizes multidimensional self-trapped states in the
form of quantum droplets~\cite{Petrov2016, Astrakharchik2018, Cabrera2018,
Cheiney2018, Semeghini2018, Hammond2022, Mithun2020}. With the LHY
correction, such states are prone to thermal and dynamical instabilities~%
\cite{Wang20201, Wang20202, Gangwar2022, Gangwar2023}.

The higher-order nonlinearity, represented by quintic terms in GPEs,
significantly affects various characteristics of the BEC ground state. In
the framework of the mean-field theory, the quintic terms naturally
represent three-body interactions~\cite{threebody1, threebody2, threebody3},
although the applicability of this setting is limited by three-body losses~%
\cite{Ketterle}. Earlier studies of BEC models that incorporate quintic
repulsive terms~\cite{Gammal2000, Bulgac2002, Tan2008, Petrov2014, Pan2021,
Hu2021} have demonstrated that they may effectively control instabilities,
leading to the formation of robust self-bounded bosonic droplets, known as
\textit{boselets}, without the need for a trapping potential~\cite%
{Bulgac2002}. Additionally, SOC also stabilizes 1D Townes solitons formed by
quintic self-attraction~\cite{Townes} against the critical collapse~\cite%
{CNSNS2, CNSNS1}. Based on these insights, we propose a model aimed at
stabilizing various quantum phases, in particular the stripe wave phase in
SOC BECs which otherwise is dynamically unstable, by incorporating
three-body interactions which are represented by quintic terms in the GPEs.
Moreover, a stable supersolid phase, created by nonlinear excitations of SOC
BECs has yet to be identified, with limited studies conducted on achieving
such a stable phase. In this paper, we demonstrate the emergence of both a
stable superfluid \textit{boselets} and a supersolid phase, incorporating
repulsive three-body interactions (R3BI) alongside attractive two-body
interactions (A2BI), into the BEC setting.

In this work, we present a robust framework for the stabilization of
quasi-1D quantum phases, specifically plane waves (PWs) and stripe waves
(SWs), in the binary SOC BECs by means of R3BI. In the absence of R3BI,
while A2BI is present, the baseband modulational instability (BBMI) and
zero-momentum-gain MI drive the formation of deterministic
rogue waves (RWs) in the PW phase~\cite{Baronio2014, Baronio2015, Liu2023},
while passband MI (PBMI) and mixedband MI (MBMI) in the SW phase give rise
to nonlinear oscillatory waves~\cite{Baronio2014,Baronio2015}. We
demonstrate that the quintic terms are responsible for the transition from
the PW phase to a stable boselet configuration, where, in particular, the
elimination of BBMI allows for the emergence of stable breathers.
Furthermore, we find that, ultimately, the suppression of instabilities in
the SW phase, characterized by stable lattice-like phonon-roton minima,
leads to the supersolid behavior~\cite{crystal1, crystal2}. These stable
quantum phases may be promising candidates for the experimental
investigation, with potential applications to spin-based quantum
simulations, spin transport, topological insulators, the quantum spin-Hall
effect, superconductivity, spintronics, and supersolids~\cite{Zutic2004,
Hasan2010, Xiao2010, Bloch2012, Amin2016, Cui2019}.

The paper is structured as follows. In Sec.~\ref{sec:2}, we introduce the
settings based on a system of mean-field coupled Gross-Pitaevskii equations
(CGPEs) with the SOC terms and two- and three-body interactions. Stability
analysis of the system, based on computation of the BdG excitation spectrum,
is presented in Sec.~\ref{sec:3}.
%The stability in the absence of the Rabi coupling is figured out in Sec.~\ref{sec:4}.
Changes in the (in)stability for the particular set of interactions are
investigated in Sec.~\ref{sec:5}. In Sec.~\ref{sec:6}, we outline the
proposal for the experimental demonstration of the predicted states of the
quantum matter. The paper is concluded in Sec.~\ref{sec:7}.

%%%%%%%%%%%%%%%%%%%%%%%%%%%%%%%%%%%%%%%%%%%%%%%%%%%%%%%%%%%%%%%%%%%%%%

\section{The mean-field model and governing dynamical equations}

\label{sec:2}
%%%%%%%%%%%%%%%%%%%%%%%%%%%%%%%%%%%%%%%%%%%%%%%%%%%%%%%%%%%%%%%%%%%%%%
We consider binary BECs with equal Rashba and Dresselhaus SOC terms and the
linear Rabi coupling between the components, assuming that a strong
transverse confinement is imposed by a cigar-shaped trapping potential \cite%
{Lewenstein}. The mean-field energy functional for spin-orbit-coupled BECs
incorporating the three-body interaction is $E=\int_{-\infty }^{+\infty }d^3
x\mathcal{E}$, with the energy density~\cite{Achilleos2013-bs, Petrov2014,
Danshita2014, Danshita2015}
\begin{align}
\mathcal{E}=\Psi ^{\dagger }H_{\mathrm{sp}}\Psi +H_{\mathrm{2B}}+H_{\mathrm{%
3B}}.  \label{eqn:source}
\end{align}%
Here, $H_{\mathrm{sp}}$ is the single-particle Hamiltonian that includes SOC
and Rabi-coupling terms, while $H_{\mathrm{2B}}$ and $H_{\mathrm{3B}}$ are
the two-body and three-body nonlinear interaction parts of the total
Hamiltonian, respectively. The particular terms in the Hamiltonian are
expressed as follows:
\begin{subequations}
\begin{align}
H_{\mathrm{sp}}& =\frac{\mathbf{p}^{2}}{2m}+V(r)+\frac{\hbar k_{L}}{m}%
p_{x}\Sigma _{z}+\hbar \Omega \Sigma _{x},  \notag  \label{eqn:sp1} \\
H_{\mathrm{2B}}& =\frac{1}{2}\bigg[g_{\uparrow \uparrow }\lvert \psi
_{\uparrow }\rvert ^{4}+g_{\downarrow \downarrow }\lvert \psi _{\downarrow
}\rvert ^{4}+2g_{\uparrow \downarrow }\lvert \psi _{\uparrow }\rvert
^{2}\lvert \psi _{\downarrow }\rvert ^{2}\bigg],  \notag \\
H_{\mathrm{3B}}& =\frac{1}{3}\bigg[\chi _{\uparrow \uparrow }\lvert \psi
_{\uparrow }\rvert ^{6}+\chi _{\downarrow \downarrow }\lvert \psi
_{\downarrow }\rvert ^{6}+3\chi _{\uparrow \downarrow }\lvert \psi
_{\uparrow }\rvert ^{2}\lvert \psi _{\downarrow }\rvert ^{2}(\psi _{\uparrow
}\rvert ^{2}\lvert +\psi _{\downarrow }\rvert ^{2})\bigg].  \notag
\end{align}%
Here, $\Psi =(\psi _{\uparrow },\psi _{\downarrow })^{T}$ is the pseudospin
two-component wavefunction, $\mathbf{p}=-\mathrm{i}\hbar (\partial
_{x},\partial _{y},\partial _{z})$ is the momentum operator, and $\Sigma
_{x,y,z} $ are, respectively, the $x$, $y$, and $z$ components of the $2\times 2$
Pauli spin matrices. The atomic mass is $m$, while $k_{L}$ and $\Omega $
denote the strengths of SOC and Rabi coupling, respectively. The tapping
potential is $V(\mathrm{r})$, and intra- and interspecies two-body
interaction strengths are $g_{\sigma \sigma }=4\pi \hbar
^{2}Na_{\sigma\sigma}/m$ and $g_{\sigma \bar{\sigma}}=4\pi \hbar
^{2}Na_{\sigma \bar{\sigma}}/m$, respectively, where $a_{\sigma \sigma }$
and $a_{\sigma \bar{\sigma}}$ are the corresponding $s$-wave scattering
lengths, where $\sigma =\uparrow ,\downarrow $ denotes the \textquotedblleft
up\textquotedblright\ and \textquotedblleft down\textquotedblright\
components, and $\bar{\sigma}$ is the opposite state of $\sigma $. Further,
the three-body interaction parameters are defined as $\chi _{\sigma \sigma
}=\lambda _{3}^{\sigma \sigma }N^{2}$ and $\chi _{\sigma \bar{\sigma}%
}=\lambda _{3}^{\sigma \bar{\sigma}}N^{2}$, with $\lambda_3 ^{\sigma \sigma
} $ and $\lambda_3 ^{\sigma \bar{\sigma}}$ representing the intra- and
interspecies three-body coupling coefficients respectively, which are
functions of the complex scattering hypervolume $D$, whose real and
imaginary parts determine the coupling constant and three-body loss,
respectively~\cite{Tan2008, Hu2021}.

To derive the quasi-1D dynamical equations, we assume that the condensate is
confined by a strong axially symmetric transverse trap ($\omega _{\perp }\gg
\omega _{x}$). Consequently, Eq.~(\ref{eqn:source}) reduces to the quasi-1D
dimensionless form through the transformation~\cite{Muruganandam2009,
Ravisankar2021cpc} $x=a_{\perp }\bar{x}$, $y=a_{\perp }\bar{y}$, $z=a_{\perp
}\bar{z}$, $t=\bar{t}/\omega _{\perp }$, and $\psi _{\uparrow ,\downarrow
}(x,y,z,t)=\bar{\psi}_{\uparrow ,\downarrow }(x,t)\bar{\psi}_{\uparrow
,\downarrow }(y,z)a_{\perp }^{-3/2}$, where
\end{subequations}
\begin{align}
\bar{\psi}_{\sigma }(y,z)=A_{y}A_{z}\exp \left( -\frac{(\omega
_{y}y^{2}+\omega _{z}z^{2})}{2}-\frac{\mathrm{i}(\omega _{y}+\omega _{z})t}{2%
}\right) ,
\end{align}%
with $A_{x}=(\omega _{y}/2\pi )^{1/4}$, and $A_{z}=(\omega _{z}/2\pi )^{1/4}$%
. By inserting above equation in Eq.~(\ref{eqn:source}), integrating over $y$
and $z$, and omitting the bar, we obtain the dimensionless form of the
CGPEs:
\begin{widetext}
\begin{subequations}
%\label{eq:gpsoc2app}
\label{eq:gpsoc:1}
\begin{align}
\mathrm{i}\partial _{t}\psi _{\uparrow }=& \bigg[-\frac{1}{2}\partial
_{x}^{2}-\mathrm{i}k_{L}\partial _{x}+V(x)+g\lvert \psi _{\uparrow }\rvert ^{2}+g_{\uparrow \downarrow }\lvert \psi _{\downarrow }\rvert ^{2}  +\chi \lvert \psi _{\uparrow }\rvert ^{4}+\chi _{\uparrow
\downarrow }\lvert \psi _{\downarrow }\rvert ^{2}\left( \lvert
\psi _{\downarrow }\rvert ^{2}+2\lvert \psi _{\uparrow }\rvert ^{2}\right) %
\bigg]\psi _{\uparrow }+\Omega \psi _{\downarrow },  \\
\mathrm{i}\partial _{t}\psi _{\downarrow }=& \bigg[-\frac{1}{2}\partial
_{x}^{2}+\mathrm{i}k_{L}\partial _{x}+V(x)+g_{\downarrow
\uparrow }\lvert \psi _{\uparrow }\rvert ^{2}+g\lvert
\psi _{\downarrow }\rvert ^{2}  +\chi \lvert \psi _{\downarrow }\rvert ^{4}+\chi _{\uparrow
\downarrow }\lvert \psi _{\uparrow }\rvert ^{2}\left( \lvert \psi
_{\uparrow }\rvert ^{2}+2\lvert \psi _{\downarrow }\rvert ^{2}\right) \bigg]%
\psi _{\downarrow }+\Omega \psi _{\uparrow }.
\end{align}%
\end{subequations}
\end{widetext}
The normalized dimensionless quantities are defined by rescaling, $%
k_{L}\rightarrow k_{L}a_{\perp }$, $\Omega \rightarrow \Omega /\omega
_{\perp }$, $g=g_{\sigma \sigma }=2a_{\sigma \sigma }N/a_{\perp }$, $%
g_{\sigma \bar{\sigma}}=2a_{\sigma \bar{\sigma}}N/a_{\perp }$, $\chi =\chi
_{\sigma \sigma }=\lambda _{3}^{\sigma \sigma }N^{2}/(3\pi ^{2}\omega
_{\perp }a_{\perp }^{6})$ and $\chi _{\sigma \bar{\sigma}}=\lambda
_{3}^{\sigma \bar{\sigma}}N^{2}/(3\pi ^{2}\omega _{\perp }a_{\perp }^{6})$.
Here, $a_{\perp }=\sqrt{\hbar /(m\omega _{\perp })}$ is the
harmonic-oscillator strength, and $V(x)=\nu ^{2}x^{2}/2$, with the trap
aspect ratio $\nu =\omega _{x}/\omega _{\perp }$, and $\omega _{\perp }=%
\sqrt{\omega _{y}\omega _{z}}$. In this paper, the dimensionless variables
are used.

In the next section, we investigate the stability of the quantum phases,
namely, PWs and SWs under the action of two- and three-body nonlinear
interactions, using the BdG analysis. This approach suggests possibilities
for the suppression of MI in these quantum phases.

%%%%%%%%%%%%%%%%%%%%%%%%%%%%%%%%%%%%%%%%%%%%%%%%%%%%%%%%%%%%%%%%%%%%%%

\section{The collective-excitation spectrum in the presence of two- and
three-body interactions}

\label{sec:3}
%%%%%%%%%%%%%%%%%%%%%%%%%%%%%%%%%%%%%%%%%%%%%%%%%%%%%%%%%%%%%%%%%%%%%%
In this section, we present the collective excitation spectrum of the
interacting SOC binary BECs for the trapless case but with the Rabi
coupling. Utilizing the BdG theory, we identify various MI types and
elaborate possibility for suppress MI, aiming to achieve stable superfluid
quantum phases.

\subsection{The stability analysis using the BdG theory}

The stability of states produced by Eq.~(\ref{eq:gpsoc:1}) is investigated
using the BdG equations derived from the linearization of Eq.~(\ref%
{eq:gpsoc:1}). We assume the conservation of the total density, $%
n=n_{\uparrow }+n_{\downarrow }$, and equal chemical potentials $\mu $ of
both components, due to the presence of the linear coupling. In the absence
of the trapping potential, $V(x)=0$, the perturbed uniform wave function is
expressed as~\cite{Goldstein1997, Abad2013},
\begin{equation*}
\psi _{\sigma }=e^{-\mathrm{i}\mu t}\left( \sqrt{n_{\sigma }}+\delta \psi
_{\sigma }\right) ,
\end{equation*}%
where ${n_{\sigma }}$ are the uniform densities of the components, and small
perturbations are introduced as
\begin{equation*}
\delta \psi _{\sigma }=u_{\sigma }\mathrm{e}^{\mathrm{i}(kx-\omega
t)}+v_{\sigma }^{\ast }\mathrm{e}^{-\mathrm{i}(kx-\omega ^{\ast }t)},
\end{equation*}%
with $u_{\sigma }$ and $v_{\sigma }$ being the perturbation amplitudes, $k$
the real wavenumber, and $\omega $ the eigenfrequency, which may be complex.
The presence of $\mathrm{Im}(\omega )\neq 0$ indicates the presence of MI.
We obtained the eigenvalue spectrum through the energy minimization,
defining the Rabi-coupling strength as $\Omega =\Omega e^{\mathrm{i}\theta }$%
, which yields $\Omega =-\Omega $ for $\theta =\pi $ ~\cite{Abad2013,
Ravisankar}. Substituting the perturbed wave functions in Eq. (\ref%
{eq:gpsoc:1}) yields
\begin{align}
(\mathcal{L}-\omega \mathrm{I})(u_{\uparrow },v_{\uparrow },u_{\downarrow
},v_{\downarrow })^{T}=0,  \label{BdG}
\end{align}%
\begin{align}
\mathcal{L}\equiv \left(
\begin{array}{cccc}
H_{\uparrow }^{+} & M_{\uparrow } & R_{\uparrow }-\Omega & R_{\uparrow } \\
-M_{\uparrow } & -H_{\uparrow }^{-} & -R_{\uparrow } & -R_{\uparrow }+\Omega
\\
R_{\downarrow }-\Omega & R_{\downarrow } & H_{\downarrow }^{-} &
M_{\downarrow } \\
-R_{\downarrow } & -R_{\downarrow }+\Omega & -M_{\downarrow } &
-H_{\downarrow }{+}%
\end{array}%
\right) ,  \label{eq:gpsoc:4}
\end{align}%
where $\mathrm{I}$ is the $4\times 4$ unit matrix, and $\mathcal{L}$ is the
BdG matrix, with
\begin{align}
H_{\sigma }^{\pm }& =\frac{k^{2}}{2}\pm k_{L}k+2gn_{\sigma }+g_{\uparrow
\downarrow }n_{\bar{\sigma}}+3\chi n_{\sigma }^{2}+\chi _{\uparrow
\downarrow }n_{\bar{\sigma}}^{2}  \notag \\
& +4\chi _{\uparrow \downarrow }n_{\sigma }n_{\bar{\sigma}}-\mu _{\sigma }
\notag \\
M_{\sigma }& =(g+2\chi n_{\sigma }+2\chi _{\uparrow \downarrow }n_{\bar{%
\sigma}})n_{\sigma }  \notag \\
R_{\sigma }& =(g_{\uparrow \downarrow }+2\chi _{\uparrow \downarrow }n_{\bar{%
\sigma}}+2\chi _{\uparrow \downarrow }n_{\sigma })\sqrt{n_{\sigma }n_{\bar{%
\sigma}}}  \notag \\
\mu _{\sigma }& =gn_{\sigma }+g_{\uparrow \downarrow }n_{\bar{\sigma}}+\chi
n_{\sigma }^{2}+\chi _{\uparrow \downarrow }n_{\bar{\sigma}}^{2}  \notag \\
& +2\chi _{\uparrow \downarrow }n_{\sigma }n_{\bar{\sigma}}-\Omega \sqrt{{n_{%
\bar{\sigma}}}/{n_{\sigma }}}.
\end{align}%
Equation~(\ref{BdG}) leads to the eigenvalue equation $\omega ^{4}-\Lambda
\omega ^{2}+\Delta =0$ with solutions
%$\omega _{\pm }^{2}=(\Lambda \pm \mathrm{i}\sqrt{4\Delta -\Lambda ^{2}})/2$,
\begin{align}
\omega _{\pm }^{2}=(\Lambda \pm \mathrm{i}\sqrt{4\Delta -\Lambda ^{2}})/2,
\label{omega}
\end{align}%
for the underlying symmetric uniform states, with $n_{\uparrow
}=n_{\downarrow }=1/2$, in the cases of $g=g_{\downarrow \uparrow }$ and $%
\chi =\chi _{\downarrow \uparrow }$. Here, we define
\begin{align}
\Lambda & =(k^{4}/2+\Omega )(k^{4}/2+X+\Omega )+k^{2}k_{L}^{2}+\Omega
(\Omega -Y)  \notag \\
\Delta & = \left[(k^{2}/2+2\Omega )(k^{2}/2+X-Y)-k^{2}k_{L}^{2}+2\Omega (Y-X)%
\right]  \notag \\
& \times \left\lbrack (k^{2}/2+2\Omega
)(k^{2}/2+X+Y)-k^{2}k_{L}^{2}\right\rbrack,
\end{align}%
with
\begin{align}
X\equiv g+\chi +\chi _{\uparrow \downarrow },~Y\equiv g_{\uparrow \downarrow
}+2\chi _{\uparrow \downarrow }.  \label{XY}
\end{align}

In the next subsection, using Eq.~(\ref{omega}), we explore various MI\
types in the presence of A2BIs and further analyze methods to eliminate them
for achieving stable quantum phases.
%%%%%%%%%%%%%%%%%%%%%%%%%%%%%%%%%%%%%%%%%%%%%%%%%%%%%%%%%%%%
\begin{figure*}[tbh]
\centering\includegraphics[width=0.98\linewidth]{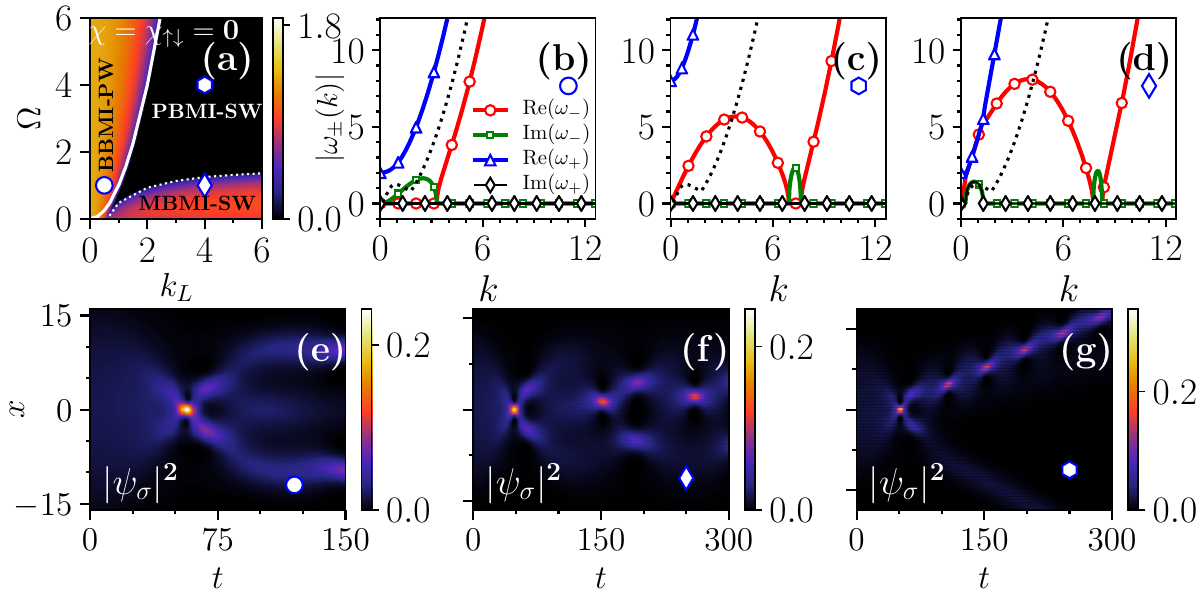}
\caption{Panel (a) shows characteristics of MI gain ($G_{-}$) in the $\left(
k_{L},\Omega \right) $ plane for fixed A2BI strengths, $g=g_{\uparrow
\downarrow }=-2$, with the perturbation wavenumber $k=1$ and R3BI strengths $%
\protect\chi =\protect\chi _{\uparrow \downarrow }=0$. The solid line
denotes the phase-transition boundary ($\Omega =k_{L}^{2}$) between the PW
and SW phases, while the dotted line indicates the transition from MBMI to
PBMI as $\Omega $ varies in the SW phase. In panels (b-d) eigenspectra show
different types of MI, \textit{viz}., BBMI, PBMI, and MBMI, respectively,
each exhibiting a distinct nonlinear dynamical behavior. The spectra
correspond to the phase plot in Fig.~\protect\ref{fig1a}(a), as indicated by
identical markers. Solid lines and symbols denote the analytical and
numerical results, respectively. Black dotted lines indicate the Feynman
dispersion derived from the structure factor. For a dense superfluid, the
Feynman energy (black-dotted lines) matches the excitation energy only in
the phonon limit ($k\rightarrow 0$); beyond this limit, the two energies
diverge. Panels (e-g) illustrate the evolution of ground-state densities ($%
\lvert \protect\psi _{\protect\sigma }\rvert ^{2}$) corresponding to the
points marked in Fig.~\protect\ref{fig1a}(a). Under effective attractive
interactions, with $X+Y<0$, the unstable phases are monitored by simulating
the evolution of the ground-state solution after quenching the interactions
and gradually ramping the trap down from $t=0$ till $t=20$.
In the
present work, we consider symmetric
inputs,
resulting in outputs which are also
symmetric with respect to the
spin-up
and spin-down components. Therefore, we
present the evolution
plots for
the single component.}
\label{fig1a}
\end{figure*}
%%%%%%%%%%%%%%%%%%%%%%%%%%%%%%%%%%%%%%%%%%%%%%%%%%%%%%%%%%%%

%%%%%%%%%%%%%%%%%%%%%%%%%%%%%%%%%%%%%%%%%%%%%%%%%%%%%%%%%%%%

\subsection{Different MI phases in the presence of attractive two-body
interactions}

%%%%%%%%%%%%%%%%%%%%%%%%%%%%%%%%%%%%%%%%%%%%%%%%%%%%%%%%%%%%
From Eq.~(\ref{omega}), we calculate the MI gain, $G_{\pm }=\lvert \mathrm{Im%
}(\omega _{\pm })|$. Scalar and spinor BECs generally show MI driven by A2BI
\cite{Kasamatsu2004, Kasamatsu2006}, while SOC and Rabi coupling can
significantly alter this behavior. We also aim to explore how R3BI terms can
affect MI of SOC binary BECs.

%\textit{The stability phase diagram} --
To investigate the stability of the perturbed SOC BECs in the $\left(
k_{L},\Omega \right) $ plane, we consider the MI gain $G_{-}$ with $%
g=g_{\uparrow \downarrow }=-2$. In this case, the self-attractive SOC BECs
exhibit three distinct types of MI in the absence of R3BI, as shown in Fig.~%
\ref{fig1a}(a). Specifically, we identify the following MI species: (i) The
baseband MI (BBMI), characterized by $\lvert \mbox{Im}(\omega )\rvert \neq 0$
at $\lvert k\rvert >0$ and $\lvert \mbox{Im}(\omega )\rvert =0$ at $k=0$,
which appears in the PW state at $\Omega \geq k_{L}^{2}$, as illustrated in
Fig.~\ref{fig1a}(b). (ii) The passband MI (PBMI), characterized by $\lvert %
\mbox{Im}(\omega )\rvert \neq 0$ at
\begin{equation*}
\lvert k\rvert >\lvert k_{\mbox{min}}\rvert >0,
\end{equation*}%
separated from $k=0$, with a gain growth starting from $k_{\mbox{min}}\neq 0$%
, as depicted in Fig.~\ref{fig1a}(c). (iii) The mixedband-MI (MBMI), a
combination of BBMI and PBMI with conditions $\lvert k_{\mbox{BBMI}}\rvert
>0 $ and $\lvert k_{\mbox{PBMI}}\rvert >\lvert k_{\mbox{min}}\rvert $, as
shown in Fig.~\ref{fig1a}(d).

In addition, we reinforce the analytical findings related to the excitation
spectrum by numerically solving the BdG equations, which also allow us to
derive the eigenvectors as a function of the wave vector $k$. To begin this
process, we consider a grid in real space that spans the range of $%
[-1000:1000]$, utilizing a step size of $\delta x=0.05$. This choice of the
grid size ensures a detailed mapping of the physical system under the
consideration. Following this, we apply the Fourier collocation method~\cite%
{yang2010nonlinear, Canuto2007spectral}, where we numerically execute the
Fourier transformation of the BdG equations. This procedure results in a
truncated reduced BdG matrix, which encompasses the essential features of
the system. We then proceed to diagonalize this matrix using the LAPACK
package~\cite{Anderson1999}, which is renowned for its efficiency in
handling such computational tasks. In terms of the momentum space, we focus
on the modes in the range of $[-50:50]$ in the $k$ direction, employing a
grid step size of $\delta k=0.0628$. This carefully chosen momentum space
grid allows us to achieve an accurate representation of the system's
behavior in the momentum space.

Following the BdG analysis of MI, we proceed to demonstrate the dynamical
(in)stability by numerically solving the full GPEs system,
%(cf. Refs. \cite{Muruganandam2009, Ravisankar2021cpc}),
aiming to obtain the ground states, using Gaussian profiles as a seed
wavefunction. Here, we present the dynamical results produced by the
numerical solution of CGPEs~(\ref{eq:gpsoc:1}) for SOC BECs. First, we
determine the ground states using the imaginary-time propagation (ITP)
method~\cite{Muruganandam2009, Ravisankar2021cpc}. Subsequently, we evolve
the ground state wavefunction through the real-time propagation (RTP). For
both ITP and RTP, we have employed the split-step Crank-Nicholson scheme~%
\cite{Muruganandam2009, Ravisankar2021cpc}. In this work, we used the grid
size of [$-250,250$] with a spacial step of $\Delta x=0.025$, and time step
of $\Delta t=10^{-4}$ for ITP and $\Delta t=5\times \Delta x^{2}$ for RTP.

Further, we follow the procedure detailed in Refs.~\cite%
{solitontrain-exp1,brightsoliton-exp2}, where the ground state was first
produced by means of ITP and subsequently quenched in RTP by switching the
cubic terms in the GPEs from repulsion to attraction. Simultaneously, the
trapping potential is gradually removed, so that $V(x)=0$ at $t\geq 20$.
Under such conditions, in the absence of R3BI and for $\Omega \geq k_{L}^{2}$%
, we observe the emergence of deterministic RWs due to the
effect of BBMI in the initial PW phase. Notably, the cubic attraction
induces RW-like dynamics~\cite{Bludov2009, Tan2022, Siovitz2023}, leading to
abrupt localization of BEC with a large amplitude, which then fragments into
two soliton trains that eventually decays, as shown in Fig.~\ref{fig1a}(e).
The emergence of an RW-like feature at $t=57$ is evident in Fig.~\ref{fig1a}%
(e), where the density attains a maximum value $\approx 0.266$, which is $15$
times higher than the initial density, $\lvert \psi _{\sigma }\rvert
_{t=0}^{2}=0.018$, which is consistent with previous findings~\cite%
{Bludov2009}. The initial exponential growth indicates the onset of the
instability, cf. Refs.~\cite{Tanzi2019, Chomaz2018}. Overall, we find that
BBMI in the attractive SOC BECs does not inherently lead to the emergence of
RWs and soliton trains, as observed in other trapped systems \cite%
{Bludov2009}.
Additionally, considering the BBMI phase and evolving the
ground state with attractive interactions under the trapping potential,
we observe the emergence of chaotic spatiotemporal patterns. This complex evolution
suggests a transition towards a regime resembling the soliton turbulence, cf.
Ref.~\cite{Sun2023, ravisankar2025csf}.

%%%%%%%%%%%%%%%%%%%%%%%%%%%%%%%%%%%%%%%%%%%%%%%%%%%%%
\begin{figure*}[tbh]
\centering\includegraphics[width=0.98\linewidth]{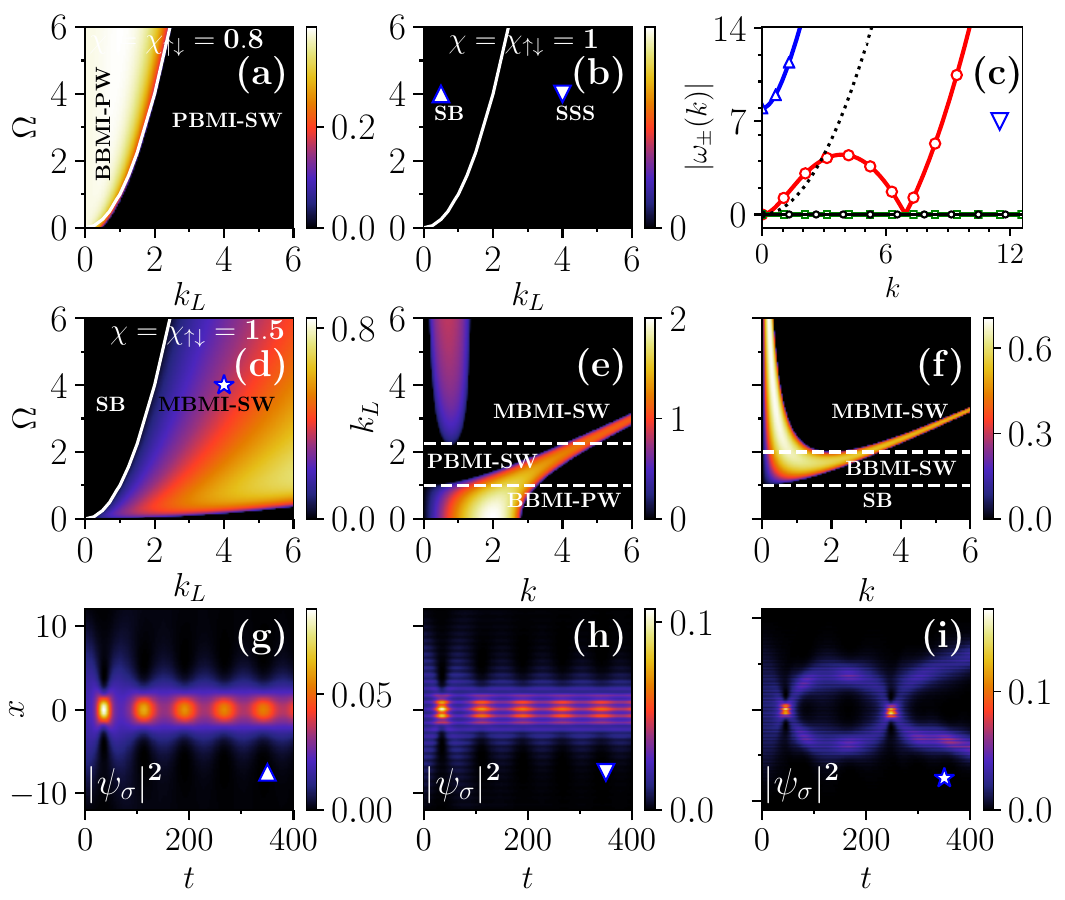}
\caption{Panels (a-c) illustrate the nature of MI in the $\left(
k_{L},\Omega \right) $ plane for fixed A2BI strengths $g=g_{\uparrow
\downarrow }=-2$, with perturbation wavenumber $k=1$ and varying R3BI
strengths: (a) $\protect\chi =\protect\chi _{\uparrow \downarrow }=0.8$, and
(b) $\protect\chi =\protect\chi _{\uparrow \downarrow }=1$. Panel (c)
depicts the excitation spectrum of the stable supersolid phase in (b),
showing a roton-phonon lattice-like state for the balanced effective
interactions $X+Y=0$, which stabilizes the unstable supersolid phase. Panel
(d) presents the instabilities for\ the effective repulsive interactions,
with $X+Y>0$, including the three-body interactions, with $\protect\chi =%
\protect\chi _{\uparrow \downarrow }=1.5$. The solid line denotes the
phase-transition boundary ($\Omega =k_{L}^{2}$) between the PW and SW
phases. For fixed $\Omega =1$, panels (e, f) show the emergence of distinct
MI gain bands for cases Fig.~\protect\ref{fig1a}(a) and Fig.~\protect\ref%
{fig1}(d), with the dashed line distinguishing different MI regimes. Panels
(g-l) illustrate the evolution of ground-state densities ($\lvert \protect%
\psi _{\protect\sigma }\rvert ^{2}$) corresponding to the points marked in
(b,c). When the system's effective interaction is modified to $X+Y\geq 0$,
the dynamics of the ground state is excited by imprinting a periodic density
modulation with wavenumber $k$, cf. Ref. \protect\cite{Astrakharchik2018}.}
\label{fig1}
\end{figure*}
%%%%%%%%%%%%%%%%%%%%%%%%%%%%%%%%%%%%%%%%%%%%%%%%%%%%%

However, in the regime with $k_{L}^{2}>\Omega $, both PBMI and MBMI occur in
the SW phase. Specifically, for relatively small $\Omega <\mathcal{R}$,
where $\mathcal{R}\equiv a-(b+a)c/(c+k_{L})$, with $a=1.6$, $b=2.7$, and $%
c=0.35 $, the system exhibits MBMI. For $k_{L}^{2}>\Omega \gtrsim \mathcal{R}
$, PBMI is observed, as shown in Fig.~\ref{fig1a}(a) (the dotted line).
Notably, PBMI is characterized by a gap, while MBMI exhibits a gapless
instability-avoided-crossing (IAC) region between $\omega _{-}$ and $%
\omega_{+} $. In the MBMI phase, $\omega _{+}$ mode is particularly
sensitive to perturbations, resulting in $G_{+}\neq 0$ and leading to RWs
accompanied by nonlinear oscillations (cf. Ref.~\cite{Baronio2014}), as
shown in Fig.~\ref{fig1a}(f). Furthermore, PBMI is associated with the
formation of breathers that propagate in $\pm ~x$ directions, as displayed
in Fig.~\ref{fig1a}(g).

Here, we identify distinct types of MI and their corresponding dynamical
behaviors, which are characterized by their instability bands, as exhibited
by the eigenvalue spectra. The respective spectra are displayed in Fig.~\ref%
{fig1a} (b-c), demonstrate excellent agreement between the analytical and
numerical results.

%%%%%%%%%%%%%%%%%%%%%%%%%%%%%%%%%%%%%%%%%%%%%%%%%%%%%%%%%%%%

\subsection{The effect of the three-body interaction on MI}

%%%%%%%%%%%%%%%%%%%%%%%%%%%%%%%%%%%%%%%%%%%%%%%%%%%%%%%%%%%%
In this subsection, we examine the impact of three-body interactions on the
system's stability, gradually increasing their strength while maintaining
the two-body attractive interaction at a fixed level.

Adding the quintic R3BI suppresses MI, exhibiting the transformation of MBMI
into weaker PBMI, as illustrated in Fig.~\ref{fig1}(a). When R3BI is
introduced under condition $X+Y<0$, with $g=g_{\uparrow \downarrow }=-2$ and
$\chi =\chi _{\uparrow \downarrow }=0.8$ [see Eq.(\ref{XY})], both BBMI and
PBMI emerge. The PBMI is further characterized by a roton instability,
defined as a well pronounced dip in the excitation spectrum with $\mathrm{Im}%
(\omega _{-})\neq 0$. The roton minimum can lead to a stable supersolid if $%
\mathrm{Im}(\omega _{-})=0$. As R3BI strength increases, the instability
associated with BBMI and PBMI becomes attenuated, approaching the PW-to-SW
transition line ($\Omega =k_{L}^{2}$), where R3BI inhibits MI. Further
enhancement of R3BI fully eliminates MI for the balanced interaction, under
condition $X+Y=0$, with fixed A2BI strengths $g=g_{\uparrow \downarrow }=-2$
and R3BI coefficient $\chi =\chi _{\uparrow \downarrow }=1.0$ [see Eq.~(\ref%
{XY})]. Thus, both the PW and SW phases exhibit stable phonon and
lattice-like phonon-roton modes with $\mathrm{Im}(\omega )=0$, as shown in
the phase plot depicted in Fig.~\ref{fig1}(b) and the respective stable
supersolid excitation spectrum in Fig.~\ref{fig1}(c). These modes are
responsible for the formation of stable boselet and supersolid phases,
respectively. However, for the repulsive effective interaction, with $X+Y>0$%
, the stability is preserved only in the PW phase, while the SW one is
subject to MBMI, see Fig.~\ref{fig1}(d).

% %%%%%%%%%%%%%%%%%%%%%%%%%%%%%%%%%%%%%%%%%%%%%%%%%%%
% \begin{figure*}[tbp]
% \centering\includegraphics[width=0.9\linewidth]{fig3-prr.pdf}
% \caption{The top row: Static density structure factors for the BBMI, PBMI,
% MBMI, and stable supersolid as depicted in Figs.~\protect\ref{fig1a} and
% \protect\ref{fig1}. The bottom row: Spin static structure factors
% corresponding to each respective mode. The dashed-dotted magenta line
% represents the Feynman criterion, corresponding to the upper branch of the
% spectrum.}
% \label{fig2}
% \end{figure*}
% %%%%%%%%%%%%%%%%%%%%%%%%%%%%%%%%%%%%%%%%%%%%%%%%%%%

We further explore MI in the $\left( k,k_{L}\right) $ plane at a fixed Rabi
strength $\Omega =1$, highlighting distinct behaviors for $\chi
=\chi_{\uparrow \downarrow }=0$ and $2$, as illustrated in Figs.~\ref{fig1}%
(e) and~\ref{fig1}(f), respectively. Initially, for $X+Y<0$, both the PW and
SW phases exhibit MI, leading to a series of novel findings. In particular,
for $\Omega \geq k_{L}^{2}$, the BBMI is present in the PW phase for $0<k<3$%
, extending to $k_{L}^{2}=\Omega $. The PBMI emerges only in the interval of
$\Omega <k_{L}^{2}\lesssim 2\Omega $, while, beyond this range, only MBMI is
observed [see Fig.~\ref{fig1}(e)]. For $X+Y>0$, comparing Fig.~\ref{fig1}(f)
and (e) yield several key insights: (i) the disappearance of BBMI and
formation of stable boselets; (ii) the transition from PBMI to BBMI; (iii)
the persistence of MBMI's characteristics.

%%%%%%%%%%%%%%%%%%%%%%%%%%%%%%%%%%%%%%%%%%%%%%%%%%%
\begin{figure*}[tbp]
\centering\includegraphics[width=0.9\linewidth]{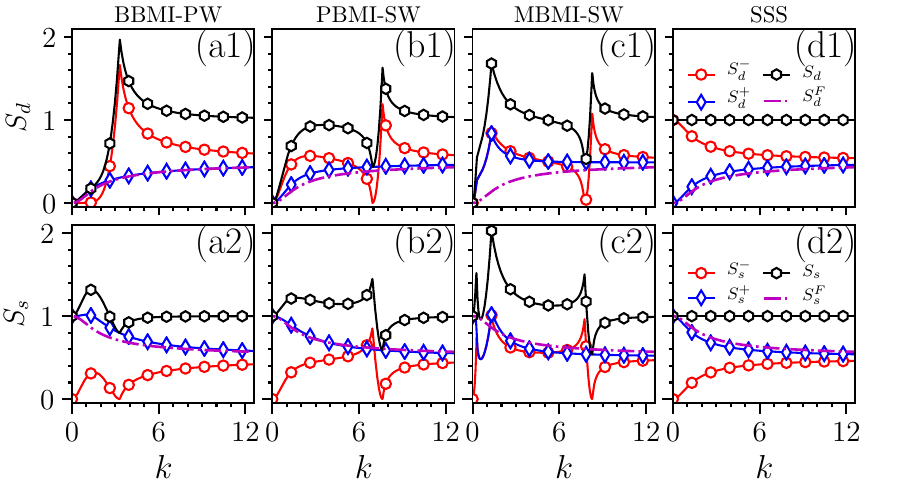}
\caption{The top row: Static density structure factors for the BBMI, PBMI,
MBMI, and stable supersolid as depicted in Figs.~\protect\ref{fig1a} and
\protect\ref{fig1}. The bottom row: Spin static structure factors
corresponding to each respective mode. The dashed-dotted magenta line
represents the Feynman criterion, corresponding to the upper branch of the
spectrum.}
\label{fig2}
\end{figure*}
%%%%%%%%%%%%%%%%%%%%%%%%%%%%%%%%%%%%%%%%%%%%%%%%%%%

In the presence of R3BI with coefficients $\chi =\chi _{\uparrow \downarrow
}=1$, the stable PW phase resembles a superfluid boselet, exhibiting stable
breather dynamics~\cite{Astrakharchik2018}, as illustrated in Fig.~\ref{fig1}%
(g). Conversely, the SW phase displays a lattice-like phonon-roton minimum
softening, which is attributable to the effect of R3BI [see Fig.~\ref{fig1}%
(c)]. The rotons are considered as a \textquotedblleft soft
mode\textquotedblright , signaling the system's approach to crystallization
into a supersolid phase~\cite{Li2013, Martone2013, JRLi2017, Liao2018,
crystal1, crystal2}. This mode facilitates the establishment of the stable
supersolid phase, see Fig.~\ref{fig1}(h). The excitation spectrum reveals
the emergence of a roton minimum without instability [$\mathrm{Im}(\omega
_{-})=0$], indicating the appearance of a stable supersolid in the SOC BECs~%
\cite{Saccani2012, Geier2021, Geier2023, Recati2023, Bland2022}. Our
analysis confirms that the system stays in a dense superfluid phase, rather
than a gaseous one, as validated by the consideration of the Feynman energy~%
\cite{Pitaevskii2016book}. For $X+Y>0$, a stable boselet forms, while the
MBMI in the supersolid phase induces nonlinear oscillations and divergence
in the $\pm x$ directions, as shown in Fig.~\ref{fig1}(i). In addition to
these MI scenarios, we observe one with the MI gain at the zero wavenumber,
characterized by $\lvert \mbox{Im}(\omega )\rvert _{k\rightarrow 0}\neq 0$,
a phenomenon that was absent in previously considered scenarios where $%
\lvert \mbox{Im}(\omega )\rvert _{k\rightarrow 0}=0$ took place. In this
case, the maximum gain occurs at $k=0$, while the minimum gain is observed
at $\lvert k\rvert <k_{\mbox{max}}$, cf. Refs.~\cite{Baronio2014, Liu2023}.
This specific MI gain arises exclusively from the strong interspecies
interaction under the condition of $X-Y+2\Omega <0$, precipitating the
emergence of deterministic RWs.

% %%%%%%%%%%%%%%%%%%%%%%%%%%%%%%%%%%%%%%%%%%%%%%%%%%%%%%%%
% \begin{figure*}[tbp]
% \centering\includegraphics[width=0.9\linewidth]{fig4-prr.pdf}
% \caption{The variation in the loss of the MI magnitude ($A_{L}^{\pm }$) and
% bandwidth ($B_{w}^{\pm }$), along with the gap between $\protect\omega _{\pm
% }$ modes ($\Delta _{g}$), is shown as a function of the interaction
% strengths for the PW (a, c) and SW (b, d) phases, respectively. In the PW
% phase, the coupling parameters are fixed to $k_{L}=0.5$ and $\Omega =1$, while
% for the SW phase the parameters are $k_{L}=4$ and $\Omega =1$.}
% \label{fig3}
% \end{figure*}
% %%%%%%%%%%%%%%%%%%%%%%%%%%%%%%%%%%%%%%%%%%%%%%%%%%%%%%%%%

%%%%%%%%%%%%%%%%%%%%%%%%%%%%%%%%%%%%%%%%%%%%%%%%%%%%%%%%%%%%%%%%%%%%%%

\subsection{Modified MI phases without Rabi coupling $\Omega =0$}

\label{sec:4}
%%%%%%%%%%%%%%%%%%%%%%%%%%%%%%%%%%%%%%%%%%%%%%%%%%%%%%%%%%%%%%%%%%%%%%

The previously analyzed nature of MI changes significantly when Rabi
coupling is set to $\Omega =0$. For $\chi =\chi _{\uparrow \downarrow }=0$,
BBMI-PW emerges in both eigenspectra $\omega_{\pm }$, in contrast, for $%
\Omega \neq 0$ $\omega_{+}$ remains stable. At $\Omega =0$, MBMI-SW exhibits
IAC behavior starting from $k=0$, while for $\Omega \neq 0$, mixedband and
passband MIs emerge. These MIs are suppressed by quintic terms for $\chi
=\chi _{\uparrow \downarrow }=1$. It was found that two gapless Goldstone
modes, along with roton-phonon lattice-like states, contribute to the
emergence of the supersolid phase. For $\chi =\chi _{\uparrow \downarrow }>1$%
, the stable boselet transforms into BBMI, while MBMI exhibits PBMI-roton
instability, all maintaining a gapless nature with $G_{+}=0$. This fact
emphasizes the importance of $\Omega $ in determining the nature of MI in
the $\omega _{+}$ state. Therefore, the physical mechanism for the stable
supersolid created in the SOC BECs~\cite{Sachdeva2020} significantly differs
from dipolar BEC, where density modulations in the stable supersolid phase
arise from the dipolar interactions linked to two gapless Goldstone modes~%
\cite{dipolar-supersolid}. However, the modified excitation spectrum,
influenced by coefficients $k_{L}$ and $\Omega $, with the 2BI and 3BI
terms, underscores the emergence of a stable supersolid in our framework.
While a stable supersolid is a hallmark of certain types of quantum matter,
including superfluid helium~\cite{helium1, helium2} and ultracold atomic
gases~\cite{Putra2020}, their stability depends on specific conditions, such
as interaction strength, trapping potentials, and temperature.

We further examine the static density and spin structure factors, defined as
$S_{d}(k)$=$N^{-1}\lvert \sum_{\sigma }\sqrt{n_{\sigma }}(u_{\sigma}(k)+v_{%
\sigma }(k))\rvert ^{2}$ and $S_{s}(k)$=$N^{-1}\lvert \sum_{\sigma}\sqrt{%
n_{\sigma }}\mbox{sgn}(\sigma)(u_{\sigma }(k)+v_{\sigma }(k))\rvert ^{2}$,
with $\mbox{sgn($\uparrow$)}$=$-\mbox{sgn($\downarrow$)}$=$1$~\cite{Abad2013}%
, revealing the system's susceptibility to density and spin fluctuations.
When $G_{\pm}\neq 0$, the $\omega _{\pm}$ branch predominantly carries spin
excitations within the corresponding MI range in $k$. In contrast, when $%
G_{\pm}=0$, the $\omega _{\pm}$ branch denotes only a density mode~\cite%
{Pitaevskii2016book}. Notably, both $S_{d}^{\pm }$ and $S_{d}$ vanish as $%
k\rightarrow 0$, where $S_{d}^{\pm }$ and $S_{s}^{\pm}$ indicate the density
of the $\omega _{\pm }$ modes and spin static structure factors,
respectively; however, for $k\neq 0$, they obey the Feynman relation, $%
S_{d}^{F}(k)=k^{2}n_{\sigma }/2\omega N$, modified for the spin structure
factor as $S_{s}^{F}(k)=1-k^{2}n_{\sigma }/2\omega N$~\cite{Liao20131,
Liao20132}.

In Fig.~\ref{fig2}(a), we show $S_{d,s}^{\pm }$ for the BBMI in the PW
phase. Here, $S_{d}^{-}$ increases monotonically across the MI range in $k$,
eventually approaching $n_{\sigma }$, while $S_{d}^{+}$ follows the Feynman
relation, as shown by the dashed-dotted magenta line. Furthermore, $%
S_{s}^{-} $ reflects spin fluctuations driven by MI, while $S_{s}^{+}$
remains unaffected by perturbations. However total $S_{d,s}$ generates
fluctuations within the MI range in the $k$ space. While PBMI also exhibits
similar fluctuations in $S_{d,s}$, the MI range of $k$ varies, as
illustrated in Fig.~\ref{fig2}(b). Compared to BBMI, PBMI maintains
stability at small values of $k$, leading to steady $S_{d,s}^{\pm }$,
whereas BBMI shows significant fluctuations. Notably, PBMI induces a
significant change in $S_{d,s}^{-}$ in the course of the onset of the roton
instability ($k\approx 8 $), indicating the dynamical instability of the
supersolid phase. In contrast, $S_{d,s}^{+}$ remains fluctuation-free,
showing no MI in the $\omega _{+}$ mode. However, Fig.~\ref{fig2}(c) reveals
unexpected fluctuations in $S_{s,d}^{+}$, confirming the emergence of MI in $%
\omega _{+} $ and the presence of IAC, where $S_{s,d}^{\pm }$ overlap within
the MI range. In stable phases, $S_{s,d}(k)=1$ with $S_{s,d}^{\pm }$ obeying
the Feynman relations [see Fig.~\ref{fig2}(d)]. In all scenarios, the
structure factors exhibit the inversion symmetry, $S(-k)=S(k)$, and increase
with momentum $k$, asymptotically approaching the limit value $S^{\pm
}(k\rightarrow \infty )=n_{\sigma }$. Hence, the total structure factor
reaches $S(k\rightarrow \infty )=2n_{\sigma }=1$, as expected.

We have explored the various types of MIs concerning the coupling
parameters, specifically PW and SW quantum phases. Additionally, we
identified conditions for obtaining stable boselets and supersolids and
investigated the role of Rabi coupling. We have analyzed the sensitivity of
the density and spin structure factors to fluctuations and their
characteristics. In the next section, we will further examine the role of
nonlinear interactions on MI.
%%%%%%%%%%%%%%%%%%%%%%%%%%%%%%%%%%%%%%%%%%%%%%%%%%%%%%%%%%%%%%%%%%%%%%

\section{The effect of interactions on MI}

\label{sec:5}
%%%%%%%%%%%%%%%%%%%%%%%%%%%%%%%%%%%%%%%%%%%%%%%%%%%%%%%%%%%%%%%%%%%%%%
In this section, we examine MIs, considering nonlinear interactions with
fixed coupling strengths. This is a crucial step of the analysis because, in
coupled BECs, intra- and interspecies interactions play a significant role
in determining the stability of the nonlinear matter-wave dynamics~\cite%
{Kasamatsu2004, Kasamatsu2006, Ravisankar2020}.

\subsection{The role of two-body and intraspecies interactions}

\begin{figure*}[tbp]
\centering\includegraphics[width=0.9\linewidth]{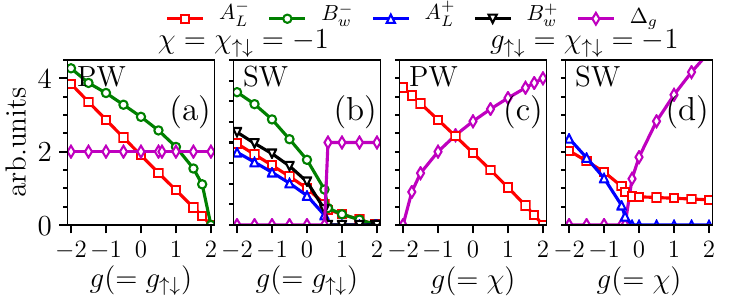}
\caption{The variation in the loss of the MI magnitude ($A_{L}^{\pm }$) and
bandwidth ($B_{w}^{\pm }$), along with the gap between $\protect\omega _{\pm
}$ modes ($\Delta _{g}$), is shown as a function of the interaction
strengths for the PW (a, c) and SW (b, d) phases, respectively. In the PW
phase, the coupling parameters are fixed to $k_{L}=0.5$ and $\Omega =1$,
while for the SW phase the parameters are $k_{L}=4$ and $\Omega =1$.}
\label{fig3}
\end{figure*}

To understand how interactions affect MI, we examine changes in the MI
magnitude ($A_{L}^{\pm }$) and bandwidths ($B_{w}^{\pm }$, where $\pm$
refers the corresponding $\omega_{\pm}$ modes), which drive the system's
nonlinear dynamical behavior~\cite{Bhat2015}. Initially, we set $\chi =\chi
_{\uparrow \downarrow }=-1$ for the PW case ($\Omega =1,k_{L}=0.5$). The MI
magnitude $A_{L}^{-}$ decreases linearly for $\omega _{-}$ as the two-body
interactions $g$, and $g_{\uparrow \downarrow }$ vary simultaneously, as
shown in Fig.~\ref{fig3}(a). When $g=g_{\uparrow \downarrow }\geq 2$, MI
disappears ($A_{L}^{-}=0$), leading to the phonon-mode softening, while
bandwidth $B_{w}^{-}$ follows the same linear trend, as shown in Fig.~\ref%
{fig3}(a). In the PW case, the $\omega _{+}$ mode is not involved in MI ($%
A_{L}^{+}=B_{w}^{+}=0$). Next, setting $\chi =\chi _{\uparrow \downarrow }=1$%
, PW does not exhibit MI in the considered range of the cubic nonlinearity, $%
[-2,+2]$. Instead, it exhibits the phonon-mode softening in the $\omega _{-}$
branch, leading to the emergence of stable boselets. However, for $g<-2$, MI
emerges, and both scenarios display a constant gapped mode, $\Delta
_{g}=2\Omega $, unaffected by the interactions. We define the gap between
the $\omega _{\pm }$ modes as $\Delta _{g}$.

%%%%%%%%%%%%%%%%%%%%%%%%%%%%%%%%%%%%%%%%%%%%%%%%%%%%%%%%
%%%%%%%%%%%%%%%%%%%%%%%%%%%%%%%%%%%%%%%%%%%%%%%%%%%%%%%%%

For the SW case ($\Omega =1,k_{L}=4$) with $\chi =\chi _{\uparrow \downarrow
}=-1$, Fig.~\ref{fig3}(b) shows $A_{L}^{\pm }$ and $B_{w}^{\pm }$, revealing
MI in both $\omega _{\pm }$ modes. Initially, the $\omega _{-}$ mode
exhibits MBMI under the A2BI, while the $\omega _{+}$ mode displays PBMI. As
the 2BI changes from attractive to repulsive, the MBMI in $\omega _{-}$
transforms to PBMI and eventually stabilizes. Concurrently, the $\omega _{+}$
mode evolves from PBMI to the stability, resulting in a stable supersolid
phase. The MBMI mode features multiple instability bands. To calculate the
magnitude of instability ($A_{L}^{\pm }$) and bandwidth ($B_{w}^{\pm }$) for
MBMI, we average the values based on the number of bands appeared in the
respective mode. In this context, the MBMI mode reveals a linear decrease in
$A_{L}^{-}$ during the transition from the attractive to repulsive cubic
nonlinearity. Without the cubic nonlinearity ($g=g_{\uparrow \downarrow }=0$%
), the system remains unstable ($A_{L}^{-}\neq 0$) due to the action of the
attractive quintic nonlinearity with $\chi =\chi _{\uparrow \downarrow }=-1$%
, indicating that the attractive 3BI alone induces the instability [see Fig.~%
\ref{fig3}(b)]. Furthermore, $A_{L}^{-}$ continues to decrease with the
increase of $g$, stabilizing at $g\geq 2$. $A_{L}^{+}$ also decreases,
reaching $A_{L}^{+}=0$ at $g=0.6$, marking the disappearance of the gapless
IAC mode and transitioning to a gapped mode between the $\omega _{\pm }$
ones. The bandwidth $B_{w}^{\pm }$ shows a similar behavior. Conversely, at $%
\chi =\chi _{\uparrow \downarrow }=1$, the stabilization point shifts to $%
g\geq -2 $ for SW, while $\omega _{+}$ is not involved.

The transition from gapped to gapless modes, based on varying the
interaction strength, is crucially important for comprehending their
behavior. With fixed two- or three-body interactions, all modes remain
gapped except for the gapless IAC mode. In the gapped mode, $\omega _{+}$
maintains a constant minimum across the interaction range, as depicted in
Fig.~\ref{fig3}(a, b). The interplay between the intra- and interspecies
two- and three-body interactions leads to diverse behaviors for the gapped
modes. Fixing the interspecies interactions ($g_{\uparrow \downarrow }=\chi
_{\uparrow \downarrow }=-1$) and simultaneously varying intraspecies
interactions $g$ and $\chi $, we observe the following trends. In the PW
phase, $\Delta _{g}$ remains zero for $g<-1.81$. Beyond this threshold, $%
\Delta _{g}$ increases exponentially with $g=\chi $. As $g$ increases
further, $A_{L}^{-}$ decreases linearly up to $g<2$, beyond which no MI
occurs, and $A_{L}^{-}=0$ [see Fig.~\ref{fig3}(c))]. In contrast, in the SW
regime, $A_{L}^{\pm }$ decrease gradually; specifically, $A_{L}^{+}=0$ and $%
A_{L}^{-}\approx 1.55$ at $g=\chi =-0.4$, while $\Delta _{g}$ remains zero.
Beyond this point, no MI is observed in the $\omega _{+}$ mode, while $%
A_{L}^{-}\neq 0$, indicating consistent instability in the SW region in the
present case. Furthermore, $\Delta _{g}$ exhibits exponential growth, as
shown in Fig.~\ref{fig3}(d).
%Additional interaction effects on MI are detailed in Appendix C.

%%%%%%%%%%%%%%%%%%%%%%%%%%%%%%%%%%%%%%%%%%%%%%%%%%%%%%%%

\subsection{The effect of the three-body and interspecies interactions}

\begin{figure*}[tbp]
\centering\includegraphics[width=0.9\linewidth]{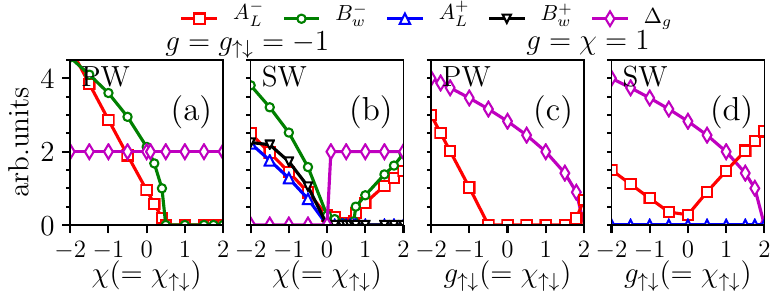}
%##\centering\includegraphics[width=0.99\linewidth]{fig3a}
\caption{The variation in the MI magnitude ($A_{L}^{\pm }$) and bandwidth ($%
B_{w}^{\pm }$), along with the gap between $\protect\omega _{\pm }$ modes ($%
\Delta _{g}$), is shown against different interaction strengths for the PW
(a, c) and SW (b, d) phases, with coupling parameters same as in Fig.~%
\protect\ref{fig3}.}
\label{fig3-supp}
\end{figure*}

%%%%%%%%%%%%%%%%%%%%%%%%%%%%%%%%%%%%%%%%%%%%%%%%%%%%%%%%
Next, we analyze the effects of the three-body interactions, with $%
g=g_{\uparrow \downarrow }=-1$. By varying 3BI strength, we attain a stable
mode for PWs at $\chi =0.5$. In comparison, $A_{L}^{-}$ and $B_{w}^{-}$ are
relatively large for $\chi <-2$, with $A_{L}^{-}$ diminishing rapidly at $%
\chi =0.5$ and entering the stable regime thereafter, as illustrated in Fig.~%
\ref{fig3-supp}(a). This fact indicates that the three-body interaction can
be employed to design dynamical patterns. For the SW ($\Omega =1,k_{L}=4$), $%
A_{L}^{-}$ and $B_{w}^{-}$ initially decrease, attaining $A_{L}^{-}=0$ at $%
\chi =0.5$. However, $A_{L}^{-}\neq 0$ holds for $\chi >0.75$, with $%
A_{L}^{-}$ increasing linearly and indicating that, with a fixed attractive
two-body interaction strengths, SWs get stabilized only for $0.5<\chi <0.75$%
, as shown in Fig.~\ref{fig3-supp}(b). The $\omega _{+}$ mode gets
stabilized and the gapless IAC disappears at $\chi =0$. It does not reappear
at $\chi >0$, suggesting the existence of a constant gapped mode.

With $g=g_{\uparrow \downarrow }=1$, the behavior of $A_{L}^{-}$ and $%
B_{w}^{-}$ for PW and SW phases is similar to that shown in Figs.~\ref%
{fig3-supp}(a) and (b). However, the PW stabilization point shifts to $\chi
=-0.5$, and SW stabilizes in interval $\chi =[-0.55,-0.5]$, with $A_{L}^{+}$
and $B_{w}^{+}$ vanishing at $\chi =-1.2$. Compared to the prior case, the
system gets stabilized with very weak repulsive two-body and attractive
three-body interactions.

%%%%%%%%%%%%%%%%%%%%%%%%%%%%%%%%%%%%%%%%%%%%%%%%%%%%%%%%
%%%%%%%%%%%%%%%%%%%%%%%%%%%%%%%%%%%%%%%%%%%%%%%%%%%%%%%%%

For the interaction parameters $g=\chi =1$, we observe a significant
transition in the PW and SW phases, from a gapped state to a gapless one, as
$g_{\uparrow \downarrow }=\chi _{\uparrow \downarrow }$ varies, see Figs.~%
\ref{fig3-supp}(c) and (d). In this interaction regime, the mode $A_{L}^{+}$
becomes zero, indicating a transition from gapless to a gapped state in the
SW phase [see Figs.~\ref{fig3}(d) and ~\ref{fig3-supp}(d)]. The stability of
the PW phase is confined to $-0.5<g_{\uparrow \downarrow }<1.8$. Outside
this range, the PW phase is unstable, as shown in Fig.~\ref{fig3-supp}(c).
The SW phase is always unstable, confirming the persistent instability under
conditions $A_{L}^{-}\neq 0$, as depicted in Fig.~\ref{fig3-supp}(d).

%%%%%%%%%%%%%%%%%%%%%%%%%%%%%%%%%%%%%%%%%%%%%%%%%%%%%%%%%%%%%%%%%%%%%

\section{Proposal for the experiment}

\label{sec:6}
%%%%%%%%%%%%%%%%%%%%%%%%%%%%%%%%%%%%%%%%%%%%%%%%%%%%%%%%%%%%%%%%%%%%%%
We further propose an experimental realization for BEC in the $^{39}$K
atomic gas \cite{Cabrera2018, Cheiney2018, Semeghini2018, Hammond2022}, with
$N\sim 10^{4}$ atoms. To create a quasi-1D cigar-shaped BEC, we consider
weak axial and strong transverse trapping frequencies: $(\omega _{x},\omega
_{\perp })/2\pi =(6,300)$\thinspace Hz, leading to a transverse length scale
of $a_{\perp }\sim 2.33\,\mathrm{\mu }\mbox{m}$. The attractive two-body
scattering lengths are $a_{\uparrow \uparrow }=a_{\downarrow \downarrow
}=a_{\uparrow \downarrow }=-4.4062~a_{0}$, where $a_{0}$ is the Bohr radius,
yielding dimensionless interaction strengths of $g=g_{\uparrow \downarrow
}\approx -2$. Such interactions can be tuned using the Feshbach resonance~%
\cite{Feshbach1, Feshbach2, Feshbach3} under the action of the magnetic
field below $507$ G~\cite{Lepoutre2016}. The two-body loss rates vanish for
symmetric two-body interactions.~\cite{Hammond2022}. For three-body
interactions, we estimate the coupling constant for R3BI as $\chi
=\{0.5-2.0\}$, corresponding to $\lambda _{3}=\{0.71-2.84\}\times 10^{-38}\,%
\mbox{m}^{6}\mbox{s}^{-1}$, which is about $\simeq 100$ times larger than
the dominant three-body loss coefficient, $K_{3}/6\sim 3\times 10^{-40}\,%
\mbox{m}^{6}\mbox{s}^{-1}$~\cite{Cheiney2018, Esry1999}. The three-body
interactions are characterized by the scattering hypervolume $D$, the
above-mentioned complex quantity whose real and imaginary components
correspond to energy shifts and three-body losses, respectively. Our results
indicate that $\left\vert \mbox{Re(D)}\right\vert \gg \left\vert \mbox{Im(D)}%
\right\vert $, placing the system near the resonance \cite{Gammal2000,
Bulgac2002, Tan2008, Pan2021}. Parameters $\Omega $ and $k_{L}$ are readily
tunable by adjusting the Raman laser frequency, wavelength, and geometry,
allowing precise control over the system's properties. In ultracold gases,
the excitation spectrum can be probed using two-photon Bragg spectroscopy
\cite{Stenger1999, StamperKurn1999, Steinhauer2002, Zwierlein2006,
Khamehchi2014, Chenshuai2015}. Therefore, our predictions are relevant for
the experimental realization.

%%%%%%%%%%%%%%%%%%%%%%%%%%%%%%%%%%%%%%%%%%%%%%%%%%%%%%%%%%%%%%%%%%%%%%

\section{Conclusions and perspectives}

\label{sec:7}
%%%%%%%%%%%%%%%%%%%%%%%%%%%%%%%%%%%%%%%%%%%%%%%%%%%%%%%%%%%%%%%%%%%%%%
%\color{blue}
We have investigated the stability of quantum phases in
spin-orbit-coupled Bose-Einstein condensates with two- and three-body
interactions, following the Bogoliubov-de-Gennes approach. Firstly, we have
found that attractive two-body interactions alone lead to various MI
(modulation instability) scenarios, including the baseband and
zero-wavenumber MIs in the PW (plane-wave) phase, which lead to the
formation of deterministic rogue waves, as seen in both scalar and vector
BEC models. Additionally, we have also identified new types of MI in the
nontrivial stripe-wave (SW) phase, resulting in passband and mixed-band MIs
that give rise to the emergence of nonlinear oscillatory matter waves and
complex matter-wave patterns. Notably, the passband MI displays the roton
instability alone, while the mixed-band MI drives phonon and roton
instabilities.

Secondly, we aimed to suppress the instabilities and achieve stable quantum
phases. In this regard, our results demonstrate that the introduction of the
three-body repulsive interactions can transform the destabilized PW phase
into a stable superfluid boselet (bosonic-droplet) phase. Additionally, we
show that the unstable SW phase can become a stable supersolid under certain
conditions. Thus, our findings reveal the emergence of supersolidity and
boselets in SW and PW phases, respectively.

Further, we have found that the density and spin structure factors, $%
S_{d,s}^{\pm }$, are sensitive to the fluctuations and characterize the
respective MI phases. As MI is present in the $\omega _{-}$ mode and absent
in the $\omega _{+}$ one, $S_{d,s}^{-}$ exhibits fluctuations, while $%
S_{d,s}^{+}$ remains in the fluctuation-free state. However, the total
structure factors $S_{d,s}$ indicate overall instability through
fluctuations. Specifically, in the MBMI phase, fluctuations in $S_{d,s}^{+}$
occur solely when an instability-avoided crossing occurs between the $\omega
_{\pm }$ modes. When a stable phase emerges, the structure factors stay
constant.

Finally, we have demonstrated the crucial role of nonlinear interactions in
achieving stable and unstable states which, further, plays an important role
in understanding the gapped nature of the spectrum. We are currently
extending the analysis to explore how repulsive three-body interactions can
lead to stable supersolid stripe phases in two-dimensional
spin-orbit-coupled BECs.

\color{black} %\section{ACKNOWLEDGMENTS}

\acknowledgments

We acknowledge financial support from NSFC under Grant No. 12174346 and No.
12374247. R.R. acknowledges the postdoctoral fellowship under Grants No.
YS304023964. H. F. acknowledges the support from Programa de Capacit\~{a}o
Institucional (CBPF) No. 31.7202/2023-5. The work of P.M. is supported by
the Ministry of Education-Rashtriya Uchchatar Shiksha Abhiyan (MoE RUSA
2.0): Bharathidasan University - Physical Sciences.
% Rashtriya Uchchatar Shiksha Abhiyan (RUSA)
The work of B.A.M. is supported, in part, by the Israel Science foundation
through grant No. 1695/22.

\twocolumngrid
%\clearpage
%\bibliographystyle{plain}
%\bibliographystyle{plain}
\bibliography{reference}

\end{document}